\documentclass[aps,pra,twocolumn]{revtex4}
\newcommand {\be}{\begin{equation}}
\newcommand {\ee}{\end{equation}}
\newcommand {\bey}{\begin{eqnarray}}
\newcommand {\eey}{\end{eqnarray}}

\usepackage{epsfig}
\usepackage{amsfonts}
\usepackage{amsmath}
\usepackage{mathbbol}
\usepackage{enumerate}
\usepackage{amsthm}
\usepackage{pgfplots}
\usepackage{tikz}

\usepackage{hyperref}

\newtheorem{definition}{Definition}
\newtheorem{theorem}{Theorem}
\theoremstyle{definition}
\newtheorem{problem}{Problem}

\newtheorem{corollary}{Corollary}

\begin{document}

\title{Optimal measurements for nonlocal correlations}
\author{Sacha Schwarz$^*$, Andr\'e Stefanov$^*$, Stefan Wolf$^\dagger$, Alberto Montina$^\dagger$}
\affiliation{$*$ Institute of Applied Physics, University of Bern, 3012 Bern, Switzerland}
\affiliation{$\dagger$ Facolt\`a di Informatica, 
Universit\`a della Svizzera italiana, 6900 Lugano, Switzerland}
\date{\today}

\begin{abstract}
A problem in quantum information theory is to find the experimental setup that maximizes 
the nonlocality of correlations with respect to some suitable measure such as the 
violation of Bell inequalities. There are however some complications with
Bell inequalities. First and foremost 
it is unfeasible to determine the whole set of Bell inequalities already for a few 
measurements and thus unfeasible to find the experimental setup maximizing their 
violation. Second, the Bell violation suffers from an ambiguity stemming from the choice 
of the normalization of the Bell coefficients. An alternative measure of nonlocality with 
a direct information-theoretic interpretation is the minimal amount of classical communication 
required for simulating nonlocal correlations. In the case of many instances simulated in 
parallel, the minimal communication cost per instance is called nonlocal capacity, and its 
computation can be reduced to a convex-optimization problem. This quantity can be computed 
for a higher number of measurements and turns out to be useful for finding the optimal 
experimental setup. Focusing on the bipartite case, we present a simple 
method for maximizing the nonlocal 
capacity over a given configuration space and, in particular, over a set of possible 
measurements, yielding the corresponding optimal setup.
Furthermore, we show that there is a functional relationship between Bell violation and 
nonlocal capacity. The method is illustrated with numerical tests and compared with
the maximization of the violation of CGLMP-type Bell inequalities on the basis of entangled 
two-qubit as well as two-qutrit states. 
Remarkably, the anomaly of nonlocality displayed by qutrits turns out to be even stronger 
if the nonlocal capacity is employed as a measure of nonlocality.
\end{abstract}

\maketitle

\section{Introduction}

A peculiarity of quantum theory is the principle of complementarity,
stating that there are complementary measurements which cannot be performed
simultaneously. Although the knowledge of the results of every possible
observation is intrinsically out of reach of experiments, complementarity 
is not \emph{per se} inconsistent with the classical realist view that results 
are intrinsic properties independent of the actual realization of a measurement.
That is, ascribing results to every possible measurement does not lead to logical 
inconsistency. However, in a seminal paper~\cite{bell}, John Bell gave a 
characterization of \emph{local realism}, implying that certain quantum correlations
cannot be explained by shared classical information under the assumption
that the measurement settings are freely chosen. Indeed, in classical terms,
such correlations can be explained by message transmission only. On the other hand,
such an influence, if it did in fact exist, would not only need to be of infinite speed
(in a preferred frame)~\cite{bancal,barnea}, but it also requires fine-tuning~\cite{wood}.

Besides their foundational importance, these \emph{nonlocal correlations} 
have gained increasing interest as an information-processing resource. For example,
they have a fundamental role in device-independent applications, such as key 
agreement in cryptography~\cite{barrett0,acin0,scarani,acin2,acin3,masanes,hanggi}
and randomness amplification~\cite{colbeck,gallego}. Furthermore, they can exponentially 
reduce the amount of communication required to solve some distributed computational 
problems~\cite{cleve,buhrman}. For some tasks, the use of nonlocal correlations can 
make communication unnecessary, such as in pseudo-telepathy games~\cite{brassard}.
Some stronger-than-quantum nonsignaling correlations can even collapse the communication 
complexity in any two-party scenario. Indeed, the access to an unlimited number of 
Popescu-Rohrlich (PR) nonlocal boxes allows two parties to solve any communication complexity 
problem with the aid of a constant amount of classical communication~\cite{vandam}.

In view of the information-processing applications of nonlocality, a practical problem 
is to find the optimal configuration of an experimental apparatus maximizing 
the strength of nonlocality. For example, one can better exploit a given
quantum state by using an optimal set of measurements.
For this purpose, it is necessary to maximize some suitable measure of nonlocality.
The set of local correlations can be characterized by a polytope whose facets are
defined by Bell inequalities~\cite{pitowski}.
Thus, the maximal violation over the whole set of inequalities can be used as a 
possible measure. However, this quantity does not have a clear information-theoretic
meaning and suffers from an ambiguity related to the way the Bell coefficients
are normalized. Furthermore, the computation of all the facets 
has generally a time cost growing more than exponentially in the number of
measurements. Indeed, a brute-force computation of every facet
has a cost growing exponentially in the number of vertices and the number
of vertices grows exponentially in the number of measurements. Although
there are better algorithms for efficiently computing the facets under some
condition~\cite{fukuda}, it is unknown whether the time complexity is actually 
polynomial. It is a fact that the whole set of Bell inequalities
has been computed only for a small number of measurements, parties, and
outcomes~\cite{collins,cabello}.
In particular, to determine the optimal experimental configuration maximizing 
the violation over the whole set of inequalities is unfeasible even for few measurements.

An alternative
measure with a more direct information-related interpretation has been employed 
in Refs.~\cite{maudlin,brassard,steiner,gisin1,gisin2,montina,bernhard} and relies on the 
very definition of nonlocality; nonlocal correlations require some communication to be 
classically simulated, thus the minimal amount of required classical communication can 
be used as a measure of the strength of nonlocality. We call this measure 
\emph{communication complexity} of the nonlocal resource. 
As shown by Pironio~\cite{pironio},
the maximal violation of the Bell inequalities and the communication complexity of 
nonlocal resources turn out to be identical if the average amount of communication 
is employed as a measure of the communication cost. However, the coefficients of each
inequality have to be rescaled by a factor whose computation requires solving 
a set of communication complexity problems. 
The recent work in Ref.~\cite{montina} mainly focused on the minimal asymptotic 
communication cost of parallel simulations in the asymptotic limit of infinite 
instances. This quantity, called {\it nonlocal capacity}, is a lower bound
on the minimal average communication cost and differs from it
by a term scaling not more than the logarithm of the 
nonlocal capacity. Thus, for high communication complexity, the two quantities 
are essentially equivalent. The {\it nonlocal capacity} is easier to be computed 
than its single-shot counterpart, considered in Ref.~\cite{pironio}. 

Importantly,
the nonlocal capacity turns out to have a functional relationship with the Bell 
violation. Namely, for every Bell inequality, the nonlocal capacity is bounded 
from below by a function of the violation. Furthermore, there is a Bell inequality 
such that the bound is tight and equal to the nonlocal capacity.
This functional relationship extends the result of Ref.~\cite{pironio} to the case
of the nonlocal capacity. 

Focusing on the bipartite case, our main goal is to introduce a method for
maximizing the nonlocal capacity over the space of experimental configurations.
For this purpose, we first need to introduce a simple algorithm for computing the 
nonlocal capacity. In Ref.~\cite{montina}, we showed that such a computation
can be reduced to a convex optimization problem, but we did not provide
an explicit numerical method. The algorithm here introduced 
is a modification of the one recently derived in
Ref.~\cite{arne} for computing the asymptotic communication complexity of 
quantum communication processes. As shown in Ref.~\cite{arne}, the algorithm
displays notable convergence properties with respect to available
optimization packages. Numerical tests with up to $27$ measurements were 
performed with a maximal computational time of the order of one hour and $6$
digits of precision on a laptop with a \emph{2.3GHz Intel Core i7} processor.
Similar convergence properties are displayed by the algorithm introduced here
for computing the nonlocal capacity. This is in 
notable contrast to the computation of all the facets of the local polytope, 
which becomes unfeasible for a much smaller size of the problem input. 
The method applies to every bipartite quantum state and, more generally, to
every nonsignaling correlation.

We then present a simple method for maximizing the nonlocal capacity over a given
configuration space and, in particular, over a set of possible measurements
with a given quantum state. The method yields the optimal experimental setup.
Furthermore, we discuss the relation between nonlocal
capacity and violation of the Bell inequalities. This relation is investigated
in numerical tests of the introduced numerical method by considering CGLMP-type 
Bell inequalities on the basis of entangled two-qubit as well as two-qutrit states. 
Remarkably, the anomaly of nonlocality displayed by entangled qutrits is even 
stronger if the nonlocal capacity is employed as a measure of nonlocality. Namely,
the maximal nonlocal capacity is exhibited for a quantum state with less 
entanglement with respect to the quantum state providing maximal violation.

The paper is organized as follows. In Sec.~\ref{sec_nonlocality}, we
introduce the \emph{nonlocal capacity} of nonsignaling correlations
and the main result of Ref.~\cite{montina}, where the computation of
the nonlocal capacity was reduced to a convex-optimization problem.
In Sec.~\ref{sec_num_algo},
we present the algorithm for computing the nonlocal capacity.
In Sec.~\ref{sec_dual_prob}, the dual form of the optimization
problem introduced in Ref.~\cite{montina} is derived. This form
is then used in Sec.~\ref{sec_opt_meas} to derive the method for maximizing 
the nonlocal capacity over a given configuration space. In particular, 
we study the optimization over the space of projective measurements.
In Sec.~\ref{bell_ineqs}, we discuss the relation between the nonlocal 
capacity and Bell inequality violation.
Finally, in Sec.~\ref{sec_test}, we present the numerical tests.

\section{Measure of nonlocal correlations}
\label{sec_nonlocality}

Here, we introduce the nonlocal capacity as a measure of
nonlocality. First, we introduce the concept of a nonsignaling box as
an abstract object producing correlated outcomes.
Then, we define the nonlocal capacity as the minimal
asymptotic communication cost required for a classical simulation
of the box. Finally, we revise the results of Ref.~\cite{montina},
where we showed that the computation of the nonlocal capacity can
be reduced to a convex-optimization problem.

\subsection{Nonsignaling boxes}
In a Bell scenario, two quantum systems are prepared in an entangled state
and delivered to two spatially separate parties, say Alice and Bob. 
Then, the parties each perform a measurement on their own system and
get an outcome. In general, Alice and Bob are allowed to choose among
their respective sets of possible measurements. We assume that the sets 
are finite, but arbitrarily large. Let us denote the measurements 
performed by Alice and Bob by the indices $a\in\{1,\dots,A\}$ and 
$b\in\{1,\dots,B\}$, respectively. 
After the measurements, Alice gets an outcome $r\in {\cal R}$ and Bob an 
outcome $s\in{\cal S}$, where $\cal R$ and $\cal S$ are two sets with
cardinality $R$ and $S$, respectively.
The overall scenario is described by the joint conditional probability 
$P(r,s|a,b)$. Since the parties are spatially separate, causality and 
relativity imply that this distribution satisfies the nonsignaling conditions
\begin{equation}
\begin{array}{c}
\label{ns-conds}
P(r|a,b)=P(r|a,\bar b)\;\; \forall a,r,b,\bar b, 
\vspace{1mm} \\
P(s|a,b)=P(s|\bar a,b)\;\; \forall b,s,a,\bar a,
\end{array}
\end{equation}
where 
$P(r|a,b)\equiv \sum_{s} P(r,s|a,b)$ and $P(s|a,b)\equiv \sum_{r} P(r,s|a,b)$
are the marginal conditional probabilities of $r$ and $s$, respectively.
In the following discussion, we consider a more general scenario including
non-quantum correlations, and we just assume that $P(r,s|a,b)$ satisfies the 
nonsignaling conditions. The abstract machine producing the correlated 
variables $r$ and $s$ from the inputs $a$ and $b$ will be called 
{\it nonsignaling box} (briefly, NS-box). 

\subsection{Nonlocal capacity}

As mentioned in the introduction, nonlocal correlations can be 
explained classically with communication between the parties. Generally,
every NS-box can be simulated through local randomness and communication.
The minimal amount of required communication is called 
\emph{communication complexity} of the NS-box. Let us denote this quantity 
by ${\cal C}_{min}$. 
NS-boxes that cannot be simulated only through mere local randomness are 
called \emph{nonlocal boxes} (NL-boxes), and their communication complexity
is strictly positive. Conversely, vanishing communication complexity is a
signature of local correlations. 
If $N$ NS-boxes are simulated in parallel, the minimal communication cost per 
instance in the limit $N\rightarrow\infty$ is called \emph{asymptotic communication 
complexity}, or \emph{nonlocal capacity}~\cite{montina}. Let us denote it by
${\cal C}_{min}^{asym}$.
Since parallel protocols are more general than single-shot protocols, we have
$$
{\cal C}_{min}^{asym}\le {\cal C}_{min}.
$$

Different definitions of communication cost can be employed, such as worst-case 
communication~\cite{brassard}, average communication~\cite{maudlin,steiner} and 
the entropy-based definition used in Ref.~\cite{montina}. The last two
are equivalent in the asymptotic case and give the same nonlocal capacity.
Hereafter, we employ the average communication for the single-shot case
in order to compare our results with some results from Ref.~\cite{pironio}.
We will consider only one-way communication from one party to the other.

\subsection{Computation of nonlocal capacity}
In Ref.~\cite{montina}, we showed that the computation of the nonlocal capacity
${\cal C}_{min}^{asym}$ is equivalent to a convex-optimization problem. 
Tight lower and upper bounds on the single-shot communication
complexity ${\cal C}_{min}$ are given in terms of ${\cal C}_{min}^{asym}$.
The optimization is made over a suitable set of probability distributions. The set,
denoted by ${\cal V}(P)$, depends on the NS-box  $P$ and is defined as
follows.
\begin{definition}
Given a NS-box $P(r,s|a,b)$, the set ${\cal V}(P)$ is defined as the set of conditional 
probabilities $\rho(r,{\bf s}|a)$ over $r$ and the sequence 
${\bf s}=\{s_1,\dots,s_{B}\}\in {\cal S}^B$ whose marginal distribution of $r$ and 
the $b$-th element of $\bf s$ is equal to $P(r,s|a,b)$.
In other words, the set ${\cal V}(P)$ contains every $\rho(r,{\bf s}|a)$ satisfying the
constraints
\begin{equation}
\label{constraints}
\sum_{{\bf s},s_b=s} \rho(r,{\bf s}|a)=P(r,s|a,b)\quad \forall a,b,r \text{ and } s,
\end{equation}
where the sum is performed over every element of the sequence $\bf s$ except the
$b$-th element $s_b$, which is set equal to $s$.
\end{definition}

The central result in Ref.~\cite{montina} is a convex-optimization problem that yields 
the nonlocal capacity of the NS-box $P$.
The nonlocal capacity is equal to the minimum of the capacity of the channels
$\rho({\bf s}|a)\equiv\sum_r \rho(r,{\bf s}|a)$ such that $\rho(r,{\bf s}|a)\in {\cal V}(P)$.
Let us recall that a channel $x\rightarrow y$ is a stochastic process defined
by a conditional probability $\rho(y|x)$ of getting the value $y$ given $x$.
Its capacity, which we denote by $C(x\rightarrow y)$, is 
the maximum of the mutual information between $x$ and $y$ over the space of 
probability distributions $\rho(x)$ of the input $x$~\cite{cover}, that is,
\begin{equation}
\label{capacity}
C(x\rightarrow y)\equiv \max_{\rho(x)} I(X;Y),
\end{equation}
the mutual information $I(X,Y)$ being defined as~\cite{cover} 
\begin{equation}
I(X;Y)=\sum_x\sum_y \rho(x,y)\log_2\frac{\rho(x,y)}{\rho(x)\rho(y)},
\end{equation}
where $\rho(x,y)$ is the joint probability distribution of $x$ and $y$,
and $\rho(x)$ and $\rho(y)$ are the marginal distributions of $x$ and $y$,
respectively. 

Given these definitions, let us introduce the functional ${\cal D}(P)$
as the minimum of the capacity $C(a\rightarrow {\bf s})$ over the distributions
$\rho(r,{\bf s}|a)\in {\cal V}(P)$.
\begin{eqnarray}
\label{cal_D}
{\cal D}(P)\equiv\min_{\rho(r,{\bf s}|a)\in {\cal V}(P)}
C(a\rightarrow {\bf s})=  \\
\nonumber
\min_{\rho(r,{\bf s}|a)\in {\cal V}(P)}\max_{\rho(a)} I(A;{\bf S}).
\end{eqnarray}

The following theorems, proven in Ref.~\cite{montina}, relate ${\cal D}(P)$
to the communication complexity and the nonlocal capacity.
\begin{theorem}
\label{main_theor}
The nonlocal capacity ${\cal C}_{min}^{asym}$ of $P$ is equal
to ${\cal D}(P)$.
\end{theorem}
\begin{theorem}
\label{theo_sshot}
The communication complexity  ${\cal C}_{min}$ is bounded by the inequalities
\begin{equation}
\label{bounds_sshot}
{\cal D}(P)\le {\cal C}_{min}\le {\cal D}(P)+2\log_2[{\cal D}(P)+1]+2\log_2 e.
\end{equation}
\end{theorem}
These inequalities hold even if the entropy-based definition of communication
is employed.
The single-shot communication complexity ${\cal C}_{min}$ is always greater than or
equal to the nonlocal capacity ${\cal C}_{min}^{asym}$. However, the difference scales 
at most logarithmically in ${\cal C}_{min}^{asym}$. 
Let us stress that the communication is from Alice to Bob and the nonlocal capacity
from Bob to Alice can take a different value.
Theorem~\ref{main_theor} reduces the computation of the nonlocal capacity to the 
following convex-optimization problem.
\begin{problem}
\label{prob1}
\begin{equation}
\begin{array}{c}
\min_{\rho(r,{\bf s}|a)} C(a\rightarrow{\bf s})  \\
\text{subject to the constraints}  \\
\rho(r,{\bf s}|a)\ge0,  \\
\sum_{{\bf s},s_b=s} \rho(r,{\bf s}|a)=P(r,s|a,b).
\end{array}
\end{equation}
\end{problem}
Note that the capacity $C(a\rightarrow {\bf s})$ is convex in $\rho(r,{\bf s}|a)$
since the mutual information is convex in $\rho(r,{\bf s}|a)$~\cite{cover} and
the pointwise maximum of a set of convex functions is a convex function~\cite{boyd}.

In general, the channel capacity does not have a known analytic expression. 
Thus, the computation of ${\cal D}(P)$ turns out to be a minimax problem 
over the variables $\rho(r,{\bf s}|a)$ and $\rho(a)$.
However, in some symmetric problems, it is possible to get rid of the 
maximization over $\rho(a)$ in Eq.~(\ref{cal_D}). This can be shown by 
using Sion's minimax theorem~\cite{minimax} 
and some general properties of the mutual information.
As the mutual information is convex in $\rho({\bf s}|a)$ and concave in 
$\rho(a)$~\cite{cover}, 
we have from the minimax theorem that the minimization and maximization 
in Eq.~(\ref{cal_D}) can be interchanged. Thus, we obtain
\begin{equation}
\label{swap_minmax}
{\cal D}(P)=\max_{\rho(a)} {\cal J}(P)
\end{equation}
where 
\begin{equation}
\label{eq_J}
{\cal J}(P)\equiv  \min_{\rho(r,{\bf s}|a)\in{\cal V}(P)} I(A;{\bf S})
\end{equation}
is a functional of $\rho(a)$. As $I(A;{\bf S})$ is concave in $\rho(a)$ and the pointwise 
minimum of a set of concave functions is concave~\cite{boyd}, the functional ${\cal J}(P)$ 
is concave. In some symmetric cases, 
it is easy to find the distribution $\rho_{max}(a)$ maximizing ${\cal J}(P)$. 
For example, if the conditional probability $P(r,s|a,b)$ is invariant under the transformation
$a\rightarrow a+1$ up to some suitable transformation of $b$, $r$ and $s$, then 
we can infer by symmetry and the concavity of ${\cal J}(P)$ that the uniform 
distribution maximizes ${\cal J}(P)$. 

Thus, if $\rho_{max}(a)$ is known, the computation of ${\cal C}_{min}^{asym}$ is reduced 
to the following convex-optimization problem.
\begin{problem}
\label{prob2}
\begin{equation}
\begin{array}{c}
\min_{\rho(r,{\bf s}|a)} I(A;{\bf S})  \\
\text{subject to the constraints}  \\
\rho(r,{\bf s}|a)\ge0,  \\
\sum_{{\bf s},s_b=s} \rho(r,{\bf s}|a)=P(r,s|a,b).
\end{array}
\end{equation}
\end{problem}

As shown later, the dual form of Problem~\ref{prob2}
is a geometric program (see Ref.~\cite{boyd}
for an introduction to dual theory). Geometric programs are an extensively
studied class of nonlinear optimization problems~\cite{gp1,gp2} and the
commercial package MOSEK (see http://www.mosek.com) provides a solver
specialized for this class. However, if the distribution $\rho_{max}(a)$
is not known and we set $\rho(a)$ equal to an arbitrary distribution, the 
solution of Problem~2 yields merely a lower bound on the nonlocal capacity.

In Sec.~\ref{sec_num_algo}, we present a simple and robust algorithm that \emph{directly}
solves Problem~\ref{prob1}. 

\section{Numerical computation of the nonlocal capacity}
\label{sec_num_algo}
The computation of the nonlocal capacity is performed through block minimization~\cite{bertsekas}.
First, let us show that the mutual information $I({\bf S};A)$ can be written
as the minimum of 
\begin{equation}
{\cal K}=\sum_{r,{\bf s},a}\rho(r,{\bf s}|a)\rho(a)\log\frac{\rho(r,{\bf s}|a)}{R(r,{\bf s}|a)}
\end{equation}
with respect to the probability distribution $R(r,{\bf s}|a)$ under the constraints 
\begin{equation}
\label{constr_R}
\begin{array}{c}
\sum_r R(r,{\bf s}|a)-\sum_r R(r,{\bf s}|\bar a)=0\;\; \forall a,\bar a,\text{ and }{\bf s},
\vspace{1mm} \\
\sum_{r,{\bf s}} R(r,{\bf s}|a)=1.
\end{array}
\end{equation}
The first equality establishes a nonsignaling condition on $R(r,{\bf s}|a)$.
The minimum is given by setting the derivative with respect to $R(r,{\bf s}|a)$ of 
the Lagrangian
\begin{equation}
\begin{array}{c}
{\cal L}={\cal K}+
\sum_a \beta(a)\left[\sum_{r,{\bf s}} R(r,{\bf s}|a)-1\right]+ 
\vspace{1mm} \\
\sum_{r,{\bf s},a,\bar a}\alpha({\bf s},a,\bar a)
\left[R(r,{\bf s}|a)-R(r,{\bf s}|\bar a)\right] 
\end{array}
\end{equation}
equal to zero, $\alpha({\bf s},a,\bar a)$ and $\beta(a)$ being Lagrange multipliers,
which are set so that the constraints~(\ref{constr_R}) are satisfied.
We obtain that the minimizer takes the form
\begin{equation}
R(r,{\bf s}|a)=\frac{\rho(r,{\bf s}|a)\rho(a)}{\beta(a)+\sum_{\bar a}\left[\alpha({\bf s},a,\bar a)-
\alpha({\bf s},\bar a,a)\right]}.
\end{equation}
The constraints are satisfied if 
$\alpha({\bf s},a,\bar a)=\frac{\rho({\bf s}|a)\rho(a)\rho(\bar a)}{\rho({\bf s})}$
and $\beta(a)=\rho(a)$, where $\rho({\bf s})\equiv\sum_a\rho({\bf s}|a)\rho(a)$. Indeed, this gives
\begin{equation}
\label{update_R}
R(r,{\bf s}|a)=\frac{\rho(r,{\bf s}|a) \rho({\bf s})}{\rho({\bf s}|a)},
\end{equation}
which trivially satisfies the constraints.
Therefore, the minimum of $\cal K$ with respect to $R(r,{\bf s}|a)$ is the mutual information
$I({\bf S};A)$. 
Thus, Eq.~(\ref{cal_D}) turns into the following minimax problem,
\begin{equation}
{\cal D}(P)=\min_{\rho(r,{\bf s}|a)\in{\cal V}(P)}\max_{\rho(a)}
\min_{R(r,{\bf s}|a)\in{\cal W}} {\cal K},
\end{equation}
where $\cal W$ is the set of nonsignaling distributions $R(r,{\bf s}|a)$ satisfying the 
constraints~(\ref{constr_R}). As $\cal K$ is linear in $\rho(a)$ and convex in $R(r,{\bf s}|a)$,
we can swap the second minimization and the maximization~\cite{minimax}, and obtain
\begin{equation}
\label{min_K0}
{\cal D}(P)=\min_{\rho(r,{\bf s}|a)\in{\cal V}(P)}\min_{R(r,{\bf s}|a)\in{\cal W}} \bar{\cal K},
\end{equation}
where
\begin{equation}
\label{funct_I0}
\bar{\cal K}\equiv \max_{\rho(a)} {\cal K}
\end{equation}
is a convex functional of $\rho(r,{\bf s}|a)$ and $R(r,{\bf s}|a)$. As done in Ref.~\cite{arne}
for the computation of the communication complexity of quantum communication processes,
a simple way to compute the nonlocal capacity is to minimize alternately
$\bar{\cal K}$ with respect to $\rho(r,{\bf s}|a)$ and $R(r,{\bf s}|a)$. This takes
to the following algorithm (see Ref.~\cite{arne} for details). \newline
{\bf Algorithm 1.} (solving Problem~\ref{prob1}). \newline
\begin{enumerate}
\item Set some initial distribution $\rho(r,{\bf s}|a)>0$.
\item\label{maxi_step}
Maximize the mutual information
\begin{equation}
\label{mutual_algo}
I({\bf S};A)=\sum_{{\bf s},a} \rho({\bf s}|a)\rho(a)\log
\frac{\rho({\bf s}|a)}{\sum_{\bar a}\rho({\bf s}|\bar a)\rho(\bar a)}
\end{equation}
with respect to $\rho(a)$ [computation of the capacity of the channel $\rho({\bf s}|a)$].
\vspace{0.5mm}
\item Set $R(r,{\bf s}|a)= \frac{\rho(r,{\bf s}|a) \rho({\bf s})}{\rho({\bf s}|a)}$
[minimization of $\cal K$ w.r.t. $R(r,{\bf s}|a)$, see Eq.~(\ref{update_R})].
\item \label{step_lambda}
Compute $\lambda(r,s,a,b)$ solving the equations
\begin{equation}
\sum_{{\bf s},s_b=s} R(r,{\bf s}|a)e^{\sum_{\bar b}\lambda(r,s_{\bar b},a,\bar b)}=P(r,s|a,b).
\end{equation}
\item\label{step_rho_sa} Set $\rho(r,{\bf s}|a)=R(r,{\bf s}|a) e^{\sum_b\lambda(r,s_b,a,b)}$
[minimization of $\cal K$ w.r.t. $\rho(r,{\bf s}|a)\in{\cal V}(P)$].
\item\label{stop_item} Stop if a given accuracy is reached (see later discussion).
\item Repeat from step~2.
\end{enumerate}
The computation at step~\ref{step_lambda} is equivalent to maximizing the functional
\begin{equation}
\begin{array}{c}
\sum_{r,s,a,b}P(r,s|a,b)\rho(a)\lambda(r,s,a,b)-  
\vspace{1mm} \\
\sum_{r,{\bf s},a}R(r,{\bf s}|a)\rho(a)e^{\sum_b\lambda(r,s_b,a,b)}
\end{array}
\end{equation}
with respect to $\lambda$~\cite{arne}, which is a convex unconstrained optimization
and can be easily done by using the Newton method. The algorithm does not 
provide only the solution of Problem~\ref{prob1}, but the computed variables 
$\lambda(r,s,a,b)$ also converge to the solution of the dual form, introduced
in the following section. 
If $\rho(a)$ maximizing ${\cal J}(P)$ is known, step~\ref{maxi_step} can be skipped and
the algorithm solves Problem~2. 

The iterations stop at step~\ref{stop_item} when a given accuracy is reached. As done 
in Ref.~\cite{arne}, the accuracy is estimated by computing the difference between 
the ${\cal K}$ and a lower bound on the nonlocal capacity derived from $\lambda(r,s,a,b)$
and $\rho(a)$. In Sec.~\ref{sec_lower_bound} we will provide a formula for computing this 
lower bound.

\section{Dual problem}
\label{sec_dual_prob}
Here, we derive the dual form of Problem~2 (See Ref.~\cite{boyd} for an introduction
to dual theory). 
The dual form of a minimization problem (primal problem) is a maximization problem whose 
maximum is always smaller than or equal to the primal
minimum, the difference being called \emph{duality gap}. However, if the constraints
of the primal problem satisfy some mild conditions such as Slater's
conditions~\cite{boyd}, then the duality gap is equal to zero. This
is the case of Problem~2. Thus, the primal and dual problems turn out to be equivalent.
As for the case of quantum communication
processes~\cite{montina2,montina3}, the dual form has some appealing properties that
make it efficient to compute lower bounds for every $P(r,s|a,b)$ given a feasible point of 
the dual constraints. These properties will be employed for the computation of the optimal
set of measurements maximizing the nonlocal correlations for a given quantum state.
Furthermore, the relationship between Bell violation and nonlocal capacity comes
directly from these properties, as shown in Sec.~\ref{bell_ineqs}.

The dual objective function is obtained by minimizing the Lagrangian with respect to the
primal variables over the domain of the primal objective function. The dual variables are 
the Lagrange multipliers associated with the primal constraints. Let us take the
set of nonnegative distributions $\rho(r,{\bf s}|a)$ as domain.
The Lagrangian of Problem~2 is 
\begin{equation}
\begin{array}{c}
{\cal L}=I({\bf S};A)- \\
\sum_{r,s,a,b}\lambda(r,s,a,b)\left[\sum_{{\bf s},s_b=s}
\rho(r,{\bf s}|a)-P(r,s|a,b)\right]\rho(a),
\end{array}
\end{equation}
which can be written in the form
\begin{equation}
\begin{array}{c}
{\cal L}=\sum_{r,s,a,b}P(r,s|a,b)\rho(a)\lambda(r,s,a,b)+ 
\vspace{1mm} \\
\sum_{r,{\bf s},a}\rho(r,{\bf s}|a)\rho(a)\big[\log\frac{\rho({\bf s}|a)}{\rho({\bf s})}
-\sum_b\lambda(r,s_b,a,b)\big]. \\
\end{array}
\end{equation}
Only the second term depends on $\rho(r,{\bf s}|a)$, and is equal to zero for 
$\rho(r,{\bf s}|a)=0$. Let us show that it is nonnegative for every distribution
$\rho(r,{\bf s}|a)$, provided that
\begin{equation}
\label{dual_constr}
\sum_a\rho(a) \max_r e^{\sum_b\lambda(r,s_b,a,b)}\le 1.
\end{equation}
The second term can be written in the form
\begin{equation}
{\cal L}_2\equiv
-\sum_{r,{\bf s},a}\rho(r,{\bf s}|a)\rho(a)\log\left[\frac{\rho({\bf s})}{\rho({\bf s}|a)} 
e^{\sum_b\lambda(r,s_b,a,b)}\right].
\end{equation}
Using Jensen's inequality and the concavity of the logarithm, we obtain
\begin{equation}
{\cal L}_2\ge-{\cal N}\log \frac{1}{\cal N}
\sum_{r,{\bf s},a}\frac{\rho(r,{\bf s}|a)\rho({\bf s})}{\rho({\bf s}|a)}
\rho(a) e^{\sum_b\lambda(r,s_b,a,b)},
\end{equation}
where ${\cal N}\equiv \sum_{\bf s}\rho({\bf s})$. This equation implies
\begin{equation}
{\cal L}_2\ge-{\cal N}\log \frac{1}{\cal N}\sum_{{\bf s},a}\rho({\bf s})
\rho(a) \max_r e^{\sum_b\lambda(r,s_b,a,b)}\ge 0,
\end{equation}
the second inequality being a consequence of Ineq.~(\ref{dual_constr}). Hence, the
minimum of ${\cal L}_2$ is equal to zero under the constraints~(\ref{dual_constr}).
Let us now
show that the minimum of ${\cal L}_2$ is $-\infty$ if Ineq.~(\ref{dual_constr})
is not satisfied for some $\bf s=\bf s'$. Let us take the distribution
\begin{equation}
\rho(r,{\bf s}|a)=\alpha \delta_{{\bf s},{\bf s}'}
\frac{\delta_{r,\bar r(a)}  e^{\sum_b\lambda(r,s_b,a,b)}}{\sum_{a'}\rho(a')\max_{r'} 
e^{\sum_b\lambda(r',s_b,a',b)}}
\end{equation}
where $\alpha$ is a positive real number and $\bar r(a)$ the maximizer of $e^{\sum_b\lambda(r,s_b',a,b)}$ 
with respect to $r$. Note that $\rho(r,{\bf s}|a)$ is not generally normalized.
Thus, the function ${\cal L}_2$ takes the form
\begin{equation}
{\cal L}_2=-\alpha\log\sum_a\rho(a) \max_r e^{\sum_b\lambda(r,s_b',a,b)}\le 0
\end{equation}
and goes to $-\infty$ for $\alpha\rightarrow+\infty$. Hence, the dual problem is the 
maximization of 
\begin{equation}
\sum_{r,s,a,b} P(r,s|a,b)\rho(a)\lambda(r,s,a,b)+I_2(\lambda)
\end{equation}
with respect to $\lambda(r,s,a,b)$, where $I_2(\lambda)$ is equal to zero if 
constraints~(\ref{dual_constr}) are satisfied and equal to $-\infty$ otherwise.
Thus, the optimization is equivalent to maximizing the objective function
\begin{equation}
\label{dual_obj_funct}
I_{dual}=\sum_{r,s,a,b} P(r,s|a,b)\rho(a)\lambda(r,s,a,b)
\end{equation}
under constraints~(\ref{dual_constr}).
Note that the constraints can also be written in the form
\begin{equation}
\sum_a\rho(a)e^{\sum_b\lambda(r_a,s_b,a,b)}\le 1 \quad \forall {\bf r}\text{ and }{\bf s},
\end{equation}
where ${\bf r}\equiv(r_1,\dots,r_A)\in{\cal R}^A$ is a sequence of elements in the set ${\cal R}$.
In this form, the optimization problem is a geometric program~\cite{gp1,gp2}.

In conclusion, the dual form of Problem~2 is 
\begin{problem}
\label{prob3}
\begin{equation}
\begin{array}{c}
\max_{\lambda} \sum_{r,s,a,b}P(r,s|a,b)\rho(a)\lambda(r,s,a,b) 
\vspace{0.5mm} \\
\text{subject to the constraints}  
\vspace{0.5mm} \\
\sum_a\rho(a) \max_r e^{\sum_b\lambda(r,s_b,a,b)}\le 1.
\end{array}
\end{equation}
\end{problem}
Performing also the maximization with respect to $\rho(a)$, we obtain the
optimization problem
\begin{problem}
\label{prob4}
\begin{equation}
\begin{array}{c}
\max_{\rho(a)}\max_{\lambda} \sum_{r,s,a,b}P(r,s|a,b)\rho(a)\lambda(r,s,a,b) 
\vspace{0.5mm} \\
\text{subject to the constraints}  
\vspace{0.5mm} \\
\sum_a\rho(a) \max_r e^{\sum_b\lambda(r,s_b,a,b)}\le 1,
\end{array}
\end{equation}
\end{problem}
which is equivalent to Problem~1. The algorithm introduced in 
Sec.~\ref{sec_num_algo} does not solve only the primal problem~\ref{prob1}, 
but computes also the Lagrange multipliers $\lambda(r,s,a,b)$ and
the distribution $\rho(a)$ solving
Problem~\ref{prob4}. The Lagrange multipliers are asymptotically approached by 
the variables computed at step~\ref{step_lambda} of the algorithm,
whereas the distribution $\rho(a)$ is approached by the variables
computed at step~\ref{maxi_step}.

The dual Problem~\ref{prob3} has some interesting properties. 
First, the objective function is linear in the input distribution $P(r,s|a,b)$  
and its computational time scales linearly in the size of the problem input, 
that is, as $R S A B$. Second, the constraints do not depend on 
the problem input $P(r,s|a,b)$. This implies that a lower bound on
the nonlocal capacity can be evaluated efficiently for every $P(r,s|a,b)$
once a feasible point of the constraints is known. These properties
will be exploited by the algorithm introduced in the next section.
Furthermore, these properties will be used in Sec.~\ref{bell_ineqs} to derive 
the functional relationship between nonlocal capacity and Bell violation.

\section{Optimizing the set of measurements}
\label{sec_opt_meas}
\subsection{General discussion}
Suppose that the conditional probability $P(r,s|a,b)\equiv P_x(r,s|a,b)$ 
depends on a parameter $x$ over some manifold and the task is to find 
the value of $x$ such that the strength of nonlocality is maximal. 
We assume that $P_x(r,s|a,b)$ is differentiable with respect to
$x$. For example,
this problem is relevant in Bell experiments for which one searches
for the optimal setup providing the highest nonlocal capacity.
This optimization method is not convex and can have
many local maxima that are not global. 
Here, we present a simple method 
for computing local maxima of the nonlocal capacity and the
associated $x$. 
The method is iterative and generates a sequence
$x_{n=1,2,\dots}$ with associated nonlocal capacity, say
${\cal C}_{n=1,2,\dots}$, which increases at each
iteration. The method employs the particular
structure of the dual Problem~3 and requires a single computation
of the nonlocal capacity plus the computation of an optimal
lower bound at each iteration, which can be done
efficiently.

Each iteration is divided in two procedures.
In the first procedure, the Lagrange multipliers $\lambda(r,s,a,b)$ and $\rho(a)$
are computed by Algorithm~1 for the value $x_n$. Then, the next value $x_{n+1}$ is 
computed by maximizing the dual objective function by keeping $\lambda(r,s,a,b)$ and 
$\rho(a)$ constant. It is worth noting that the second procedure is equivalent
to the maximization of the violation of a Bell inequality.
The general algorithm is as follows.\newline
{\bf Algorithm 2.} 
\begin{enumerate}
\item $n=1$ and set $x_1$ equal to some initial value. 
\item\label{algo2_step2}
Compute the Lagrange multipliers 
$\lambda_n$ and the distribution $\rho_n(a)$ for the conditional probability
$P_{x_n}(r,s|a,b)$. The computation is made by Algorithm~1.
\item\label{step_find_x}
Compute the maximizer $\bar x$ of
\begin{equation}
\sum_{r,s,a,b}P_x(r,s|a,b)\rho_n(a)\lambda_n(r,s,a,b)
\end{equation}
with respect to $x$ and set $x_{n+1}=\bar x$.
\item Stop if the maximization at the previous step does not
make enough progress.
\item $n=n+1$.
\item Repeat from step~\ref{algo2_step2}.
\end{enumerate}
Let us show that the sequence ${\cal C}_n$ generated by Algorithm~2 
increases monotonically.
As $P_{x_{n+1}}(r,s|a,b)$ maximizes the objective function~(\ref{dual_obj_funct})
with $\lambda=\lambda_n$ and $\rho(a)=\rho_n(a)$, we have that
$$
\begin{array}{c}
\sum_{r,s,a,b} P_{x_{n+1}}(r,s|a,b)\rho_n(a)\lambda_n(r,s,a,b)\ge 
\vspace{1mm} \\
\sum_{r,s,a,b} P_{x_n}(r,s|a,b)\rho_n(a)\lambda_n(r,s,a,b)={\cal C}_n.
\end{array}
$$
The left-hand side of the inequality provides a lower bound on the nonlocal 
capacity ${\cal C}_{n+1}$, since $\lambda_n$ and $\rho_n(a)$ are a feasible
point of the optimization Problem~4. Thus, ${\cal C}_{n+1}\ge {\cal C}_n$.
Although the nonlocal capacity increases at each iteration,
this does not guarantee that the convergence is toward a maximum.
A convergence proof of this algorithm is made difficult by the
implicit form of the nonlocal capacity as a function of $x$.
Furthermore, this function is not guaranteed to be differentiable,
even if $P(r,s|a,b)$ is differentiable.
Nonetheless, numerical simulations show that the sequence always
converges toward a local maximum. As said, the optimization
is not convex, and many trials with different initial values of 
$x$ have to be performed.

\subsection{Optimal search with a given quantum state}

Now, let us consider the specific problem of finding the optimal
quantum-measurement setup for a given fixed quantum state. Specifically,
we introduce an algorithm for solving step~\ref{step_find_x} of
Algorithm~2. The procedure is essentially equivalent to maximizing the
violation of a Bell inequality and can be used also for that
purpose. Let
$\hat\rho$ be the density operator of the two systems on which
Alice and Bob each perform a projective measurement. Let the number
of measurement outcomes $R$ and $S$ be the dimension of 
the Hilbert space associated to Alice and Bob's systems, respectively. 
Each measurement of Alice and Bob is characterized by a set of 
$R$ and $S$ orthogonal vectors, respectively, each vector being 
associated with an outcome. Let us denote the $i$-th
vector of the $m$-th measurement performed by Alice and Bob by
$|\alpha_{m,i}\rangle$ and $|\beta_{m,i}\rangle$, respectively.
The conditional probability $P(r,s|a,b)$ takes the form
\begin{equation}
P(r,s|a,b)=\langle\alpha_{a,r}|\langle\beta_{b,s}|\hat\rho |\beta_{b,s}\rangle
|\alpha_{a,r}\rangle.
\end{equation}
The objective function of the dual problem takes the form
\begin{equation}
I_{dual}=\sum_{r,s,a,b}
\langle\alpha_{a,r}|\langle\beta_{b,s}|\hat\rho |\beta_{b,s}\rangle
|\alpha_{a,r}\rangle\rho(a)\lambda(r,s,a,b).
\end{equation}
At step~3 of Algorithm~2, we have to maximize this function
with respect to the vectors $|\alpha_{a,i}\rangle$ and 
$|\beta_{b,i}\rangle$ by keeping $\lambda(r,s,a,b)$ and $\rho(a)$ constant.
The maximization is performed by keeping the orthogonality relations
among the vectors associated with the same measurement. The method
used in the optimization is not critical, as the hard part
of Algorithm~2 is the computation of the nonlocal capacity. To
find the maximum, we can use a block-maximization by alternately
maximizing with respect to the vectors $|\alpha_{a,i}\rangle$
by keeping $|\beta_{b,i}\rangle$ constant and {\it vice versa}.
Let us first consider the case of two outcomes for each
measurement, that is, the case with $R=S=2$.

\subsubsection{Two-dimensional case}
The outcomes $r$ and $s$ take two possible values, say $\pm1$.
We consider only the maximization with respect to Alice's vectors
$|\alpha_{a,r=\pm1}\rangle$, as the procedure on the other
block is identical.
The objective function is quadratic in the vectors
$|\alpha_{a,i}\rangle$ and takes the form
\begin{equation}
I_{dual}=\sum_{r,a}
\langle\alpha_{a,r}|\hat\rho_A(r,a)|\alpha_{a,r}\rangle\rho(a)
\end{equation}
where
\begin{equation}
\hat\rho_A(r,a)\equiv \sum_{s,b}
\langle\beta_{b,s}|\hat\rho |\beta_{b,s}\rangle\lambda(r,s,a,b).
\end{equation}
The maximization of $I_{dual}$ is performed with the orthogonality
constraints
\begin{equation}
\langle\alpha_{a,r}|\alpha_{a,r'}\rangle=\delta_{r,r'}.
\end{equation}
As the optimizations over vectors associated with different measurements
are decoupled, we can perform them separately. For the sake of simplicity,
let us drop the index $a$ and write the objective function as
\begin{equation}
\sum_r \langle\alpha_r|\hat\rho_A(r)|\alpha_r\rangle\equiv J
\end{equation}
The core problem is to solve an optimization problem of the form
\begin{problem}
\begin{equation}
\begin{array}{c}
\max_{|\alpha_r\rangle} J
\vspace{0.5mm} \\
\text{subject to the constraints}  
\vspace{0.5mm} \\
\langle\alpha_r|\alpha_{r'}\rangle=\delta_{r,r'}.
\end{array}
\end{equation}
\end{problem}
Let us consider the unitary matrix
\begin{equation}
\hat U(\eta)=e^{\eta|\alpha_{-1}\rangle\langle \alpha_1|-
\eta^*|\alpha_1\rangle\langle \alpha_{-1}|},
\end{equation}
where $\eta$ is a complex number, and define the
pair of orthogonal vectors 
\begin{equation}
\begin{array}{c}
|\alpha_{1},\eta\rangle=\hat U(\eta)|\alpha_1\rangle, \\
|\alpha_{-1},\eta\rangle=\hat U(\eta)|\alpha_{-1}\rangle.
\end{array}
\end{equation}
The set $\{|\alpha_{\pm1}\rangle\}$ is the optimizer of
$J$ only if 
\begin{equation}
\label{max_cond}
\left. 
\frac{d}{d\eta} \sum_r \langle\alpha_r,\eta|\hat\rho_A(r)|\alpha_r,\eta\rangle\right|_{\eta=0}
=0,
\end{equation}
the symbol $\eta$ being dealt as a independent variable with respect to
the complex conjugate $\eta^*$.  Eq.~(\ref{max_cond}) implies that
\begin{equation}
\label{optimal_2D}
\langle\alpha_1|\left[\hat\rho_A(1)-\hat\rho_A(-1)\right]|\alpha_{-1}\rangle=0.
\end{equation}
Thus, the pair of orthogonal vectors maximizing $J$ are such that 
$\hat\rho_A(1)-\hat\rho_A(-1)$
is diagonal in that basis, that is, the pair is given by the eigenvectors of 
$\hat\rho_A(1)-\hat\rho_A(-1)$. There are only two solutions. Depending on the order 
of the vectors in the pair, we have the maximum or the minimum.

\subsubsection{Higher dimensions}
In the higher-dimensional case, the most general unitary matrix takes the form
\begin{equation}
\hat U(\hat\eta)=e^{\sum_{i,j}\eta_{ij}|\alpha_i\rangle\langle \alpha_j|}.
\end{equation}
where $\hat\eta$ is a $R\times R$ anti-Hermitian matrix with elements 
$\eta_{ij}$.
Let us define the vectors
$|\alpha_r,\hat\eta\rangle\equiv \hat U(\hat\eta)|\alpha_r\rangle$.
The set $\{|\alpha_r\rangle\}$ is a stationary point of $J$ if and only if 
\begin{equation}
\label{cond_der}
\left. 
\frac{d}{d\eta_{ij}} \sum_r \langle\alpha_r,\hat\eta|\hat\rho_A(r)|\alpha_r,\hat\eta\rangle\right|_{\hat\eta=0}
=0,
\end{equation}
the symbol $\eta_{ij}$ being dealt as a independent variable with respect to
the complex conjugate $\eta_{ij}^*=-\eta_{ji}$. Condition~(\ref{cond_der}) implies 
the optimality condition
\begin{equation}
\label{cond_maxi}
\langle\alpha_i|\left[\hat\rho_A(i)-\hat\rho_A(j)\right]|\alpha_{j}\rangle=0.
\end{equation}

These equations are equivalent to the optimality condition~(\ref{optimal_2D}) of the
two-dimensional case, applied to every pair of vectors $|\alpha_r\rangle$.
To solve Eqs.~(\ref{cond_maxi}), we can maximize cyclically over every pair. This
procedure monotonically increases the functional $J$. We expect that the generated
sequence asymptotically converges toward a stationary point, as each maximization will 
make progress until the conditions~(\ref{cond_maxi}) are satisfied for 
every $i$ and $j$. Indeed, it can be shown that the convergence is implied by 
Zoutendijk's theorem~\cite{wright}. Numerical tests show that the procedure 
quickly converges toward the maximum of $J$, the computation taking a time that
is negligible with respect to the computation of the nonlocal capacity at
step~\ref{algo2_step2} of Algorithm~2. Possibly, $J$ could have local maxima
that are not global. Thus, one could need to repeat the procedure with 
different initial conditions and check if the iteration converges to
different local maxima.

Algorithms~1 and 2 are the main results of this paper. In the next section, we will
discuss the relationship between Bell violation and nonlocal capacity, which
is computed by Algorithm~1 and optimized by Algorithm~2 over a given configuration
space.

\section{Nonlocal capacity and Bell violation}
\label{bell_ineqs}
In Ref.~\cite{pironio}, Pironio proved that the minimal average amount of communication
required by a classical simulation of nonlocal correlations turns out to be equal to 
the maximal violation of the Bell inequalities, once the inequalities are suitably 
normalized. Here, we prove a similar result and show that there is a 
functional relationship between Bell violation and nonlocal capacity. Namely, given
a Bell inequality, we prove that the nonlocal capacity is bounded from below by a 
function of the violation.
Furthermore, there is an optimal Bell inequality such that the bound turns out
to be equal to the nonlocal capacity. The optimal inequality is not necessarily
a facet of the local polytope. Let us first introduce the local polytope and
the definition of Bell inequalities.

\subsection{Local polytope}
The correlations between the outcomes $r$ and $s$ associated with the measurements 
$a$ and $b$ are local if and only if the conditional probability $P(r,s|a,b)$ takes
the form
\begin{equation}
\label{local_cond}
P(r,s|a,b)=\sum_x P_A(r|a,x)P_B(s|b,x) P_S(x),
\end{equation}
where $P_A$, $P_B$, and $P_S$ are suitable probability distributions. In this
case, the correlations can be simulated through shared randomness and no 
communication is required. In particular, the nonlocal capacity is equal to 
zero if and only if the correlations are local. It is always possible to write the 
conditional probabilities $P_A$ and $P_B$ as convex combination of deterministic 
processes, that is,
\begin{equation}
\begin{array}{c}
P_A(r|a,x)=\sum_{\bf r} P_A^{det}(r|{\bf r},a) \rho_A({\bf r}|x), 
\vspace{1mm} \\
P_B(s|b,x)=\sum_{\bf s} P_B^{det}(s|{\bf s},b) \rho_B({\bf s}|x),
\end{array}
\end{equation}
where ${\bf r}\equiv(r_1,\dots,r_A)$, ${\bf s}\equiv(s_1,\dots,s_B)$, 
$P_A^{det}(r|{\bf r},a)=\delta_{r_a,r}$ and
$P_B^{det}(s|{\bf s},b)=\delta_{s_b,s}$.
Using this decomposition, Eq.~(\ref{local_cond}) takes the form of a
convex combination of deterministic distributions. That is,
\begin{equation}
P(r,s|a,b)= \sum_{{\bf r},{\bf s}} 
P_A^{det}(r|{\bf r},a) P_B^{det}(s|{\bf s},b) \rho_{AB}({\bf r},{\bf s}),
\end{equation}
where 
$\rho_{AB}({\bf r},{\bf s})=\sum_x \rho_A({\bf r}|x) \rho_B({\bf s}|x) P_S(x)$.
Thus, the set of local distributions is a polytope, called \emph{local polytope},
defined by $R^A S^B$ vertices. Each vertex is specified by the sequences ${\bf r}$ 
and $\bf s$ and is given by the deterministic distribution 
$P_A^{det}(r|{\bf r},a)P_B^{det}(s|{\bf s},b)$. Since the elements of the
local polytope are normalized distributions and satisfy the nonsignaling
conditions~(\ref{ns-conds}), the $R S A B$ parameters defining $P(r,s|a,b)$
are not independent and the polytope lives in a 
lower-dimensional subspace. The dimension of this subspace and, more 
generally, of the subspace of NS-boxes is equal to~\cite{collins}
\begin{equation}
d_{NS}\equiv A B(R-1)(S-1)+A(R-1)+B(S-1).
\end{equation}

By the Minkowski-Weyl theorem,
the local polytope can be represented as the intersection of finitely many half-spaces.
A half-space is defined by an inequality
\begin{equation}
\label{bell_ineq}
\sum_{r,s,a,b}P(r,s|a,b)B(r,s;a,b)\le L.
\end{equation}
In the case of the local polytope, these inequalities are called \emph{Bell inequalities}.
A minimal representation of a polytope is given by the set of facets of the
polytope. A half-space $\sum_{r,s,a,b}P(r,s|a,b)B(r,s;a,b)\le L$
specifies a facet if the associated hyperplane 
$\sum_{r,s,a,b}P(r,s|a,b)B(r,s;a,b)=L$ intersects the boundary of
the polytope in a set with dimension equal to the dimension of the
polytope minus one. A distribution $P(r,s|a,b)$ is local if and only if
every facet inequality is not violated. To check the violation
of every inequality is not generally a tractable problem, but to test 
the membership to the local polytope can be done in polynomial 
time~\cite{pitowski}. 

Generally, the parameter $L$ in Ineq.~(\ref{bell_ineq}) is chosen so that 
the boundary $\sum_{r,s,a,b}P(r,s|a,b)B(r,s;a,b)=L$ of the half-space
touches the local polytope, that is, so that the boundary contains at least
one vertex (see for example Ref.~\cite{pironio}). This is attained 
by taking
\begin{equation}
L=\max_{{\bf r},{\bf s}} \sum_{a,b} B(r_a,s_b;a,b).
\end{equation}
If a distribution $P(r,s|a,b)$ is nonlocal, it violates some Bell inequality
and the strength of the violation is given by the positive quantity
\begin{equation}
\Delta B\equiv \sum_{r,s,a,b}P(r,s|a,b)B(r,s;a,b)-L.
\end{equation}

The maximum of $\Delta B$ over the whole set of Bell inequalities can be
used as a measure of the violation. However, this measure suffers from an
ambiguity. Indeed, the coefficients in Eq.~(\ref{bell_ineq}) are uniquely
defined by the half-space up to a multiplicative constant. Although, the
multiplicative constant does not affect the order in the violation
strength for each inequality, this is not the case for the maximal violation 
over the whole set of inequalities, since it is possible to choose different 
constants for each inequality. A general rule
is to set the multiplicative constant so that $L$ is equal to $2$, which
is the value used in the CHSH inequality introduced in Ref.~\cite{chsh}.
However, there is another ambiguity since the local polytope has a dimension 
$d_{NS}$ lower than the number of parameters $B(r,s;a,b)$. The ambiguity 
stemming from the nonsignaling conditions does not affect the strength 
of the violation. However, this is not the case for the ambiguity associated 
with the normalization of $P(r,s|a,b)$. Indeed, 
the transformations $B(r,s;a,b)\rightarrow B(r,s;a,b)+K(a,b)$ and 
$L\rightarrow L+\sum_{a,b}K(a,b)$ do not change the half-space
in the subspace of normalized distributions, but it changes
the strength of the violation once the transformed $L$ is normalized
to $2$. Another general rule is
to fix partially the additional terms $K(a,b)$ by setting the
quantity at the left-hand side of Ineq.~(\ref{bell_ineq}) equal to
$0$ in the case of uniform distributions. Although this does not determine
uniquely $K(a,b)$, it fixes the ambiguity on the violation strength.
Besides this ambiguity, the Bell violation does not have a clear
information-theoretic meaning. In the next subsection, we introduce
a functional relation between Bell violation and nonlocal capacity.
This relation fixes the aforementioned ambiguity by providing
an information-theoretic meaning to the Bell violation.

\subsection{Lower bounds on the nonlocal capacity}
\label{sec_lower_bound}
In Sec.~\ref{sec_dual_prob}, we introduced Problem~\ref{prob4}, 
which is the dual form of Problem~\ref{prob1}. Its solution gives the
nonlocal capacity of the NS-box $P(r,s|a,b)$. As already stressed previously,
an appealing property of the dual problem is that the constraints do
not depend on $P(r,s|a,b)$. Thus, the objective function gives
a lower bound on the nonlocal capacity for every $P(r,s|a,b)$,
provided that $\rho(a)$ and $\lambda(r,s,a,b)$ satisfy the constraints. 
Furthermore, the objective function has the linear form of the left-hand
side of a Bell inequality~(\ref{bell_ineq}).

If $\rho(a)$ and $\lambda(r,s,a,b)$ do not satisfy the constraints,
a feasible point can be easily generated with the transformation
\begin{equation}
\lambda(r,s,a,b)\rightarrow \lambda(r,s,a,b)+B^{-1} K_\lambda,
\end{equation}
where $K_\lambda$ is a suitable constant. Namely, it is sufficient to set
\begin{equation}
K_\lambda=-\log\max_{\bf s}\sum_a\rho(a)\max_r e^{\sum_b\lambda(r,s_b,a,b)}.
\end{equation}
Thus, for every $\rho(a)$ and $\lambda(r,s,a,b)$, we have 
\begin{equation}
\label{bound0}
{\cal C}_{min}^{asym}\ge\sum_{r,s,a,b}P(r,s|a,b)\rho(a)\lambda(r,s,a,b)+K_\lambda.
\end{equation}
The right-hand side of this inequality provides a lower bound on
the nonlocal capacity and can be used in Algorithm~1 for computing
the accuracy at each iteration. The computed lower bound converges
to the nonlocal capacity as $\lambda(r,s,a,b)$ and $\rho(a)$ converge
to the solution of Problem~\ref{prob4}.

Every function $\rho(a)\lambda(r,s,a,b)$ can be associated with a
Bell half-space by identifying $\rho(a)\lambda(r,s,a,b)$ with the
Bell coefficients $B(r,s;a,b)$ up to a multiplicative factor. That is,
\begin{equation}
\gamma\rho(a)\lambda(r,s;a,b)\leftrightarrow B(r,s;a,b),
\end{equation}
where $\gamma$ is some constant. It is convenient to define the non-normalized
function $\eta(a)=\gamma\rho(a)$, which completely determines 
$\gamma$ and $\rho(a)$. Namely, we have $\gamma=\sum_a\eta(a)$ and 
$\rho(a)=\gamma^{-1}\eta(a)$.
In terms of the Bell coefficients $B(r,s;a,b)$ and the violation $\Delta B$, 
Ineq.~(\ref{bound0}) takes the form
\begin{equation}
\label{bound_nl_cap}
{\cal C}_{min}^{asym}\ge\frac{\Delta B+L}{\gamma}
+K_\eta.
\end{equation}
where
\begin{equation}
K_\eta\equiv -\log
\max_{\bf s} \sum_a \rho(a) \max_r e^{\eta^{-1}(a)\sum_b B(r,s_b;a,b)}.
\end{equation}
This inequality holds for every non-negative function $\eta(a)$.
For local correlations, the right-hand side of the inequality is non-positive,
since the nonlocal capacity is equal to zero. This can be directly checked
by using the Jensen inequality in the last term. Indeed, we have
\begin{equation}
\begin{array}{c}
e^{-K_\eta}=\max_{{\bf r},{\bf s}}\sum_a
\rho(a) e^{\eta^{-1}(a)\sum_bB(r_a,s_b;a,b)}
\vspace{1mm} \\
\ge e^{ \gamma^{-1}\max_{{\bf r},{\bf s}}\sum_{a,b}B(r_a,s_b;a,b)}
\end{array}
\end{equation}
which implies, by definition of $L$, that
\begin{equation}
\label{from_jensen}
K_\eta\le -\gamma^{-1} L.
\end{equation}

Note that the bound on the nonlocal capacity can be negative even if the associated Bell
inequality is violated, since the left-hand side of Ineq.~(\ref{from_jensen}) is
generally different from $-\gamma^{-1} L$. However, the difference can be made arbitrarily 
small by taking $\eta(a)$ sufficiently large. Indeed, in the limit of large $\eta(a)$,
the exponential in $K_\eta$ can be well approximated by its linear expansion.

We can get rid of $\eta(a)$ by maximizing the right-hand of Ineq.~(\ref{bound_nl_cap}) 
over the space of non-negative $\eta(a)$. We have
\begin{equation}
\label{lower_b}
{\cal C}_{min}^{asym}\ge F(\Delta B),
\end{equation}
where 
\begin{equation}
\label{def_F}
F(\Delta B)\equiv\max_{\eta(a)\ge0}\left[
\frac{\Delta B+L}{\gamma}+K_\eta\right].
\end{equation}
The function $F(\Delta B)$ has now the nice feature of being positive if and
only if the violation $\Delta B$ is positive.
Note that $F(\Delta B)$ depends on the Bell coefficients $B(r,s;a,b)$.
Every Bell inequality has an associated function $F(\Delta B)$, which provides
a lower bound on the nonlocal capacity. Furthermore, there is an optimal inequality
such that $F(\Delta B)$ is a tight bound and turns out to be equal to the nonlocal
capacity, as stated by the following. \newline
\begin{theorem}
\label{theorem_bell_c}
Given an NS-box $P(r,s|a,b)$ there is an optimal set of Bell coefficients $B(r,s;a,b)$
such that ${\cal C}_{min}^{asym}=F(\Delta B)$. The Bell coefficients are 
$B(r,s;a,b)=\rho(a)\lambda(r,s,a,b)$, where $\rho(a)$ and $\lambda(r,s;a,b)$
are solutions of Problem~\ref{prob4}.
\end{theorem}
{\it Proof.} Let $\lambda(r,s,a,b)$ and $\rho(a)$ be the solution of Problem~\ref{prob4}.
Thus, 
\begin{equation}
{\cal C}_{min}^{asym}=\sum_{r,s,a,b}P(r,s|a,b)\rho(a)\lambda(r,s,a,b).
\end{equation}
Furthermore,
\begin{equation}
\label{ineq_th}
\sum_a\rho(a)\max_r e^{\sum_b\lambda(r,s_b,a,b)}\le1. 
\end{equation}
{\it Proof.}
Let us take $B(r,s;a,b)=\rho(a)\lambda(r,s;a,b)$ and $\eta(a)=\rho(a)$. Then, 
Ineq.~(\ref{ineq_th}) implies that $K_\eta\ge0$. This inequality,
the definition of $\Delta B$ and the definition of $F(\Delta B)$ imply that
${\cal C}_{min}^{asym}\le F(\Delta B)$. As also the inequality 
${\cal C}_{min}^{asym}\ge F(\Delta B)$ holds, the theorem is proven. $\square$

\begin{corollary}
The set of quantum measurements maximizing the the nonlocal capacity maximizes also the
violation of the optimal Bell inequality defined by the coefficients
$B(r,s;a,b)=\rho(a)\lambda(r,s,a,b)$, where $\rho(a)$ and $\lambda(r,s;a,b)$
are solutions of Problem~\ref{prob4}.
\end{corollary}
This corollary is quite obvious. Indeed, if the violation of the optimal inequality
was not maximal, then step~\ref{step_find_x} of Algorithm~2 would find another
experimental setup such that the nonlocal capacity is greater, in contradiction
with the hypothesis. The corollary implies that the set of measurements
maximizing the nonlocal capacity also maximizes the violation of a facet-defining Bell 
inequality if $\rho(a)\lambda(r,s,a,b)$ are the coefficients of such a Bell inequality.

As mentioned previously, the definition of the Bell coefficients $B(r,s;a,b)$ suffers
from an ambiguity stemming from the nonsignaling conditions satisfied by
$P(r,s|a,b)$. Namely, given a real function $A(r,s;a,b)$ such that
$\sum_{r,s,a,b}P(r,s|a,b)A(r,s;a,b)=0$ for every nonsignaling $P(r,s|a,b)$,
the transformation $B(r,s;a,b)\rightarrow B(r,s;a,b)+A(r,s;a,b)$ does not 
change the Bell half-space in the subspace of nonsignaling distributions.
This ambiguity does not affect the value of the Bell violation,
but it can affect the value of the second term at the right-hand side of 
Eq.~(\ref{def_F}). The same feature is also present in the bounds derived by 
Pironio~\cite{pironio}.
The dependence of $F(\Delta B)$ on the extra-term $A(r,s;a,b)$ means
that each Bell inequality is associated with an infinity of bounds.
We can get rid of this dependence by performing a further maximization
over $A(r,s;a,b)$, so that we have the bound
\begin{equation}
\bar F(\Delta B)\equiv\max_{\eta(a)\ge0,A\in {\cal A}}\left[
\frac{\Delta B+L}{\gamma}+\bar K_\eta\right],
\end{equation}
where $\bar K_\eta$ is obtained from $K_\eta$ by replacing the coefficients
$B$ with $B+A$,
and $\cal A$ is the set of functions $A(r,s;a,b)$ such that
$\sum_{r,s,a,b}P(r,s|a,b)A(r,s;a,b)=0$ for every nonsignaling $P(r,s|a,b)$.

\section{Numerical tests}
\label{sec_test}

In this last section, we illustrate the introduced optimization method
through some numerical tests on entangled qubits as well as entangled 
qutrits. The method is compared with the maximization of the violation 
of facet-defining Bell inequalities. The considered quantum states take the form 
\begin{equation}\label{eq:StateWithGamma}
\vert \psi(\gamma_1,\gamma_2) \rangle  = 
\frac{\vert 0\rangle_A \vert 0\rangle_B + 
\gamma_1 \vert 1\rangle_A \vert 1\rangle_B + 
\gamma_2\vert 2\rangle_A \vert 2\rangle_B}{\sqrt{1+\gamma_1^2+\gamma_2^2}},
\end{equation}
with $\gamma_1 \in [0,1]$ and $\gamma_2\in\{0,1\}$. Entangled qubits and qutrits
corresponds to $\gamma_2=0$ and $\gamma_2=1$, respectively.

We first consider the case of entangled qubits ($\gamma_2=0$) with two measurements
and two outcomes, and compute numerically the set of measurements maximizing the nonlocal
capacity as well as the violation of the CHSH inequality. The resulting two optimal
sets turn out to be very similar for every $\gamma_1$ and identical for the maximally 
entangled state ($\gamma_1=1$). Namely, the set maximizing the violation of 
a facet-defining Bell inequality is approximately optimal also for the nonlocal capacity.
We also find that the Bell inequality such that $\bar F(\Delta B)$ is 
the nonlocal capacity is a facet of the local polytope for $\gamma_1=1$. 
The study is then extended to the case of qutrits, for which we maximize numerically
the violation of the Collins-Gisin-Linden-Massar-Popescu inequality~\cite{cglmp2002,schwarz2014}
(CGLMP$3$). The resulting optimal measurement setting turns out to be
notably different from the setting maximizing the nonlocal capacity in
a range of $\gamma_1$ between about $0.5$ and $0.8$.
This implies that the Bell inequality with
maximal $\bar F(\Delta B)$ is far away from being a CGLMP facet. In fact,
it turns out that the inequality is not close to any facet of the local
polytope. We also find that the anomaly of nonlocality is even stronger if the 
nonlocal capacity is employed instead of the CGLMP violation.

Thereby, we follow the notation of 
Ref. \cite{collins}, referring to the Bell inequalities for a given Bell scenario as 
Bell-$ABRS$ inequality, where $A$ and $B$ represent the number of measurements and $R$ 
and $S$ the number of outputs for Alice and Bob, respectively.
The left-hand side of the Bell inequality~(\ref{bell_ineq}) is denoted by the symbol 
$\mathcal{B}_{ABRS}$, namely,
\begin{equation}
\mathcal{B}_{ABRS} \equiv\sum_{r,s,a,b} P(r,s|a,b) B(r,s;a,b).
\end{equation}
The conditional probability $P(r,s|a,b)$ and the coefficients $B(r,s;a,b)$ will
be occasionally represented also as $RSAB$-dimensional vectors $\vec P$ and $\vec B$,
respectively.

\subsection{Clauser-Horne-Shimony-Holt Inequality}
\label{sec:CHSH}

\begin{figure}[t!]
\input{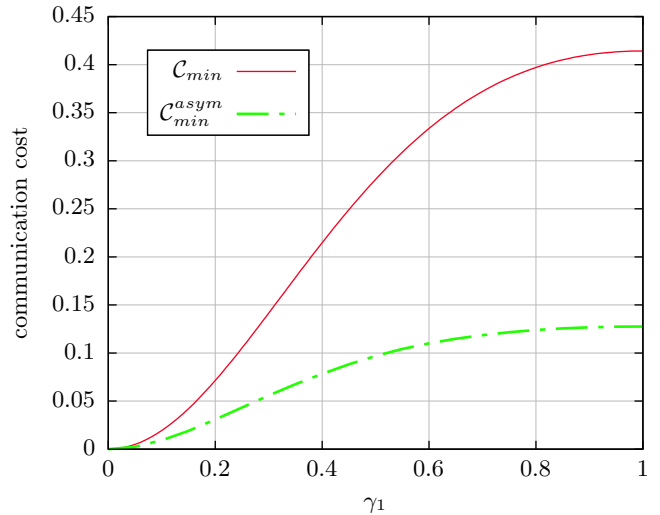}
\caption{Minimal average communication cost ${\mathcal{C}}_{min}$ (solid line) and nonlocal 
capacity $\mathcal{C}_{min}^{asym}$ as functions of $\gamma_1$ for entangled qubits
($\gamma_2=0$).}
\label{fig:I2andNLC_Qubit}
\end{figure}

In the simplest bipartite Bell scenario with two settings and two outcomes per party, the 
local polytope has dimension $8$ with $24$ facets, i.e. $16$ positivity facets and $8$ CHSH 
facets. Let the outcomes $r$ and $s$ take the values $\pm1$. Thus, a CHSH inequality 
takes the form
\begin{equation}
\begin{array}{r}
\sum_{r,s} r s \left[P(r,s|0,0)-P(r,s|0,1)+\right.
\vspace{1mm}  \\
\left. P(r,s|1,0)+P(r,s|1,1)\right]\le 2.
\end{array}
\end{equation}
The other CHSH inequalities are obtained by permuting the outcome values and exchanging
the measurement settings.

According to Ref.~\cite{pironio}, the violation of any suitably normalized  Bell inequality 
sets a lower bound on the single-shot communication complexity of a NS-box. Furthermore,
there is an optimal Bell inequality such that the violation turns out to be equal to
the communication complexity. In the Bell-$2222$ scenario, the optimal inequality
is a facet of the local polytope~\cite{pironio}. Namely, we have
\begin{equation}
{\mathcal{C}}_{min} = 
\frac{1}{2}\mathcal{B}_{2222}-1, 
\end{equation}
provided that the facet with maximal violation is taken.
We have computed the set of measurements maximizing the nonlocal capacity ${\cal C}_{min}^{asym}$ 
as well as the communication complexity ${\cal C}_{min}$, that is, the violation. 
In Fig.~\ref{fig:I2andNLC_Qubit}, we report the corresponding values of ${\cal C}_{min}^{asym}$ 
and ${\cal C}_{min}$ as functions of $\gamma_1$. The two measures display similar
behavior, although the nonlocal capacity turns out to be quite smaller than the
communication complexity. The two quantities satisfy the inequalities 
${\cal C}_{min}^{asym} \le {\mathcal{C}}_{min} 
\leq \mathcal{C}_{min}^{asym} + 2 \log(\mathcal{C}_{min}^{asym}+1)+ 2 \log_2 e$,
which come from Theorems~\ref{main_theor},\ref{theo_sshot}.

Algorithm~1, used for computing the nonlocal capacity, generates also the functions 
$\rho(a)$ and $\lambda(r,s,a,b)$ that are solutions of Problem~\ref{prob4}. Thus,
the bound $\bar F({\Delta B})$ that is maximal and equal to the nonlocal
capacity is associated with the Bell coefficients $B(r,s;a,b)=\rho(a)\lambda(r,s,a,b)$, 
as stated by Theorem~\ref{theorem_bell_c}. The coefficients maximizing the
bound are unique up to the transformation $B(r,s;a,b)\rightarrow B(r,s;a,b)+A(r,s;a,b)$,
where $A(r,s;a,b)$ is any function in $\cal A$ (defined at the end of 
Sec.~\ref{sec_lower_bound}). This transformation changes the
components of the vector $\vec B$ which are orthogonal to the local polytope.
Whereas the vector $\vec B$ defining a CHSH inequality is parallel to the local
polytope, this is not the case for the vector computed from the solution of Problem~4.
To compare the computed $B(r,s;a,b)$ with the facet-defining coefficients, we
have removed the orthogonal components by computing the projection $\vec B_\parallel$
of $\vec B$ onto the subspace of the NS-boxes. Then, we have 
evaluated the scalar product between the normalized vector $\vec B_\parallel$ and 
the normalized CHSH vector. Let us denote this quantity by
\begin{equation}
S_B\equiv\frac{{\vec B_\parallel}\cdot{\vec B_f}}{\|\vec B_\parallel\| \|\vec B_f\|},
\end{equation}
where $\vec B_f$ is the vector orthogonal to a CHSH facet.
For the maximally entangled state ($\gamma_1=1$), $S_B$ is equal to $1$ and, thus,
the coefficients $B(r,s;a,b)$ turn out to define a facet of the local polytope. 
This also implies that the measurement setup maximizing the violation is also
optimal for the nonlocal capacity. 
The quantity $S_B$ decreases by decreasing $\gamma_1$ and reaches the minimum 
$0.86$ at about $\gamma_1=0.48$. Thus, the maximal angle between $\vec B_\parallel$ 
and $\vec B_f$  is about $30$ degrees. At first glance, this angle seems to be
quite large. However, one has to keep in mind that the nonsignaling space has
dimension $8$ and two randomly generated vectors tend to be almost orthogonal
in high-dimensional spaces for the principle of the concentration of measure.
In particular, the probability that two randomly generated $8$-dimensional
vectors have an angle smaller than $30$ degrees is about $0.3\%$.

We have then compared the optimal set for the nonlocal capacity
with the set obtained by maximizing the violation of CHSH inequalities.
Namely, we have evaluated the nonlocal capacity by taking the measurement setup
maximizing the Bell violation. For every $\gamma_1$ the resulting value
differs from the maximal nonlocal capacity by a small value not greater
than about $10^{-3}$. Thus, the tilt between $\vec B_\parallel$ and 
$\vec B_f$ has a small effect on the optimal configuration, which can
be computed with good approximation by merely maximizing the violation
of the CHSH inequality.

\subsection{Collins-Gisin-Linden-Massar-Popescu Inequality}
\label{sec:CGLMP}

Let us now consider the case of entangled qutrits with two measurements and three 
outcomes per party. In this Bell-$2233$ scenario, the local polytope lies in a 
hyperplane of dimension $24$ and consists of $1116$ facets. Thereby, besides 
36 positivity facets, we encounter $432$ CGLMP3 facets as well as $648$ facets which 
can be identified as liftings of the CHSH inequality~\cite{lifting}. 
To compute the facets, we used 
the software package FAACETS~\cite{rosset2014,faacets}. Denoting by 
$P(r_a=s_b+k)$ the probability that the outcomes $r_a$ and $s_b$ of measurements 
$a$ and $b$ differ by $k$ modulo $3$, a CGLMP3 inequality
takes the form~\cite{cglmp2002}
\begin{equation}
\begin{array}{c}
P(r_0=s_0)+P(s_0=r_1+1)+P(r_1=s_1)
\vspace{1mm} \\
+P(s_1=r_0)-P(r_0=s_0-1)-P(s_0=r_1)
\vspace{1mm} \\
-P(r_1=s_1-1)-P(s_1=r_0-1)\le 2.
\end{array}
\end{equation}
The violation of this inequality, divided by $2$, gives a lower bound on
the single-shot communication complexity of a NS-box~\cite{pironio}.
This lower bound turns out to be equal to the communication complexity,
provided that the measurement setting maximizes the violation~\cite{pironio},
which is the case considered here. Thus, 
\begin{equation}
{\mathcal{C}}_{min}=\frac{1}{2}\mathcal{B}_{2233}-1.
\end{equation}
\begin{figure}[ht]
\input{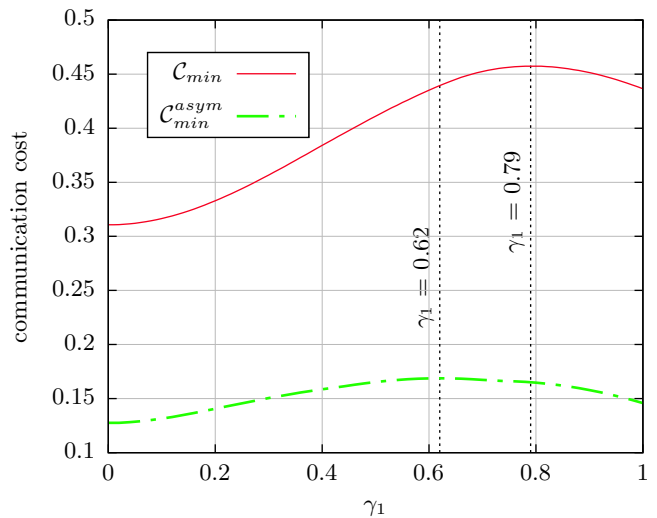}
\caption{Bell violation ${\mathcal{C}_{min}}$ (solid line) and nonlocal capacity 
$\mathcal{C}_{min}^{asym}$ (dashed line) as functions of $\gamma_1$ for entangled
qutrits ($\gamma_2=1$).}
\label{fig:I3andNLC_Qutrit}
\end{figure}
\begin{figure}[ht]
\input{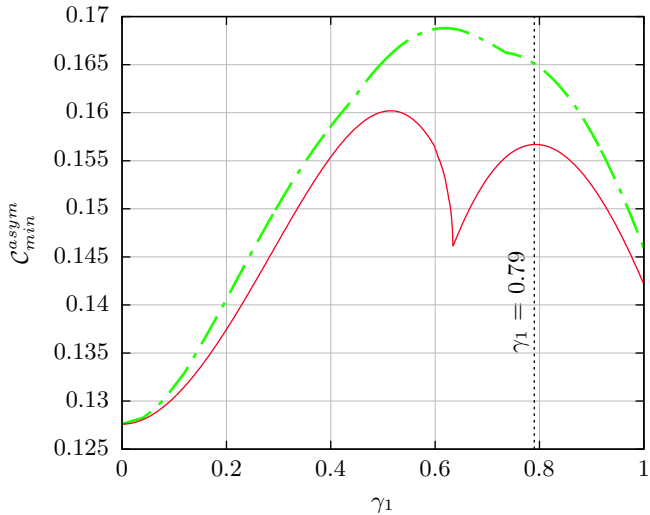}
\caption{Nonlocal capacity of entangled qutrits as a function of $\gamma_1$,
maximized over the set of measurements (dashed line), and computed by using
the measurement setting with maximal violation of the CGLMP3 inequality (solid line).}
\label{fig:NLC_withPhiAPhiB_wrtI3max}
\end{figure}

As done in the case of entangled qubits, 
we have computed the set of measurements maximizing the nonlocal capacity ${\cal C}_{min}^{asym}$ 
as well as the normalized CGLMP violation, that is,  ${\cal C}_{min}$.
In Fig.~\ref{fig:I3andNLC_Qutrit}, we report the corresponding values of ${\cal C}_{min}^{asym}$ 
and ${\cal C}_{min}$ as functions of $\gamma_1$. Also in this case, the two quantities satisfy 
the inequalities ${\cal C}_{min}^{asym} \le {\mathcal{C}}_{min} 
\leq \mathcal{C}_{min}^{asym} + 2 \log(\mathcal{C}_{min}^{asym}+1)+ 2 \log_2 e$,
although they are evaluated with optimal measurement settings that are actually notably 
different, as we will see.

In Ref.~\cite{latorre}, it was shown that the largest violation of the CGLMP inequality
is exhibited for a non-maximally entangled state, namely, for $\gamma_1\simeq 0.79$.
This behavior, known as anomaly of nonlocality~\cite{methot}, shows that there is
not a monotonic relationship between strength of entanglement and strength
of nonlocal correlations, if the latter is defined as Bell violation and two
measurements per party are considered. Remarkably, besides the fact that both 
curves in Fig.~\ref{fig:I3andNLC_Qutrit} are non-monotonic, 
the maximal value for 
the nonlocal capacity is taken at $\gamma_1\simeq 0.62$, which is significantly lower than
the value at which the Bell violation is maximal. Namely, if the nonlocal capacity is 
employed as a measure of nonlocality instead of the CGLMP violation, the maximal 
strength of nonlocality is exhibited for a quantum state with even less entanglement.

This higher anomaly is displayed by taking a set of measurements that is optimal
for the nonlocal capacity, but it becomes even stronger if the set of measurements
maximizing the Bell violation is taken. In Fig.~\ref{fig:NLC_withPhiAPhiB_wrtI3max}, 
we report the nonlocal capacity evaluated with this set (solid line) as well as 
the maximal nonlocal capacity (dashed line). The former displays two local maxima.
One maximum is at $\gamma_1\simeq 0.79$, where also the Bell violation is maximal.
The other one is the absolute maximum and is at $\gamma_1\simeq 0.5$. It is worth
to note that the optimal setup maximizing the CGLMP violation is independent of 
$\gamma_1$ for values of the parameter between about $0.63$ and $1$. The analytic
expression of this set of measurements is given in Ref.~\cite{cglmp2002}.
Below $0.63$, the maximizer becomes a function of $\gamma_1$. Curiously,
this threshold is the value at which the cusp in 
Fig.~\ref{fig:NLC_withPhiAPhiB_wrtI3max} is located.

The notable
difference between the two curves in Fig.~\ref{fig:NLC_withPhiAPhiB_wrtI3max}
implies that the measurement setup maximizing
the CGLMP violation is far away from being a good approximation of the optimal
setup for the nonlocal capacity. Thus, the two optimization methods produce
notably different optimal sets of measurements.

\section{Conclusion}

In this paper, we have presented a simple algorithm for computing the nonlocal
capacity of nonlocal correlations, which provides a measure of nonlocality as an 
alternative to the extent of violations of Bell inequalities. The algorithm is an 
adaptation of a method introduced in Ref.~\cite{arne} for quantum channels to the 
case of nonlocal correlations. Then, we have introduced an algorithm for maximizing
the nonlocal capacity with respect to the experimental setup. In particular,
we have considered the maximization with respect to the measurement setting.
The method has been applied to the case of qubits and qutrits. In the
case of qubits, the maximization of the nonlocal capacity does not produce
a measurement setting that is notably different from the optimal configuration maximizing
the CHSH violation. Conversely, in the case of non-maximally entangled qutrits, 
the two maximization methods turn out to produce notably different optimal setups.
Remarkably, the anomaly of nonlocality showed in 
Ref.~\cite{latorre} becomes even stronger once the nonlocal capacity 
is employed as a measure of nonlocality. If the set of measurements maximizing
the CGLMP violation is used, the nonlocal capacity displays two local maxima,
the absolute maximum being taken for a quantum state that is notably less 
entangled than the quantum state maximizing the CGLMP violation~\cite{latorre}.

We have also showed that, for every Bell inequality, there is a function
of the violation providing a lower bound on the nonlocal capacity. Furthermore,
there is an optimal Bell inequality such that the function turns out to
be equal to the nonlocal capacity. The optimal inequality does not necessarily
define a facet of the local polytope. This relationship between nonlocal
capacity and Bell violation is an adaptation of the results of 
Ref.~\cite{pironio} to the case of the asymptotic communication complexity.
The lower bounds on the nonlocal capacity and on the single-shot communication
complexity derived in Ref.~\cite{pironio} are essentially equivalent
in the limit of large communication complexity. This equivalence, which is
not evident by scrutinizing the mathematical expressions of the bounds, can 
be a fruitful object of future investigation.

Unlike the measure of entanglement, which is essentally unique for pure states 
and equal to the entropy of entanglement~\cite{popescu,bennett}, there are
different possible measures of nonlocality, such as the one considered
in Ref.~\cite{vandam2}, which is defined as the relative entropy $D(P||P_L)$ 
between a given joint distribution $P(r,s|a,b)$ and the closest local 
distribution $P_L(r,s|a,b)$ minimizing $D$. Interestingly, 
besides the maximization of the nonlocal capacity, the introduced method
can be applied for maximizing this other measure. For this
purpose, it is sufficient to derive the dual form of the original
minimization problem. The resulting dual objective function is identical
to the dual objective function derived here, and the dual constraints
display the same properties that we have used to derive the algorithm
optimizing the experimental setup.

Finally, the optimization problem can be used for solving an open question
concerning Werner states. A Werner state is a mixture between a maximally
entangled state and the identity density operator. In the case of entangled
qubits, the Werner state admits a local model if the probability weight of
the maximally entangled state, say $\gamma$, is smaller than $0.659$~\cite{acin4} 
and is nonlocal for $1/\sqrt2\le \gamma\le1$, as the CHSH inequalities are violated. 
In Ref.~\cite{vertesi}, V\'ertesi derived a
family of Bell inequalities that are violated for $\gamma>0.7056$, which is slightly
below the bound $1/\sqrt2$. This family requires $465$ measurement settings
on each side. Thus, the value, say $\gamma_0$, at which the transition local-nonlocal
occurs is between $0.659$ and $0.7056$.
Is it possible to derive a better upper bound on $\gamma_0$ with a much smaller
set of measurements? To answer this question, in Ref.~\cite{montina},
the nonlocal capacity was computed for a number of measurements up to $20$ by 
trying a high number of different settings, such as highly symmetric settings
and random configurations. However, we always found a transition at $\gamma=1/\sqrt2$.
The optimization algorithm introduced in this paper can help to find a
better set of measurements for which the transition occurs at a lower value
of $\gamma$. Note that the algorithm provides also the Bell inequality
that is violated.

{\it Acknowledgments.} 
This work is supported by the Swiss National Science Foundation (grant PP00P2\_133596),
the NCCR QSIT, 
the COST action on Fundamental Problems in Quantum Physics, and Hasler foundation
through the project "Information-Theoretic Analysis of Experimental Qudit Correlations".

\bibliography{biblio.bib}

\end{document}